\documentclass{aastex}
\usepackage{emulateapj5}

\begin{document}

\title{ NEW TEMPERATURES OF DIFFUSE INTERSTELLAR GAS: THERMALLY UNSTABLE GAS}

\author{Carl Heiles}

\affil {Astronomy Department, University of California,
    Berkeley, CA 94720-3411}
\email{heiles@astron.berkeley.edu}

\begin{abstract}
	We present new Arecibo 21-cm line measurements of the
temperatures of interstellar gas.  Our temperatures for the Cold Neutral
Medium (CNM) are significantly lower than previous single-dish results
and in very good accord with theoretical models.  For warm gas at $T >
500$ K, we find a significant fraction of gas ($>47\%$) to lie in the
thermally unstable region $500 \rightarrow 5000$ K; moreover, about
$60\%$ of all the neutral atomic gas has $T > 500$ K.  Large amounts of
thermally unstable gas are not allowed in theoretical models of the 
global interstellar medium. 
\end{abstract}


\section{Introduction}

          In 1977 McKee \& Ostriker (MO) extended the two-phase model of
Field, Goldsmith, \& Habing (1969) by including supernovae (SN).  Their
interstellar medium (ISM) has four phases, if one includes the Warm
Ionized Medium, and is dominated by individual SN explosions.  The Hot
Ionized Medium (HIM) fills the interior of SN remnants and powers their
blast waves, which sweep up the gas inside the bubbles and pile it into
the shells.  Soon this shocked gas starts to cool and recombine rapidly,
forming the Cold Neutral Medium (CNM).  Soft X-rays produced by adjacent
HIM penetrate the outsides of CNM clouds, heating the gas to form the
Warm Neutral Medium (WNM). 

	The swept-up gas consists of two distinct neutral phases in
physical contact, each in thermally stable equilibrium with equal
thermal pressures.  It is difficult to study such gas in individual SN
remnants, but for superbubbles the general picture of hot gas inside a
swept-out volume surrounded by dense walls is well corroborated by
modern multiwavelength studies, for example in Eridanus (Heiles,
Haffner, \& Reynolds 1999). 

	Theoretical CNM and WNM temperatures are derived by calculating
the equilibrium temperature as a function of thermal pressure.  There
exist two stable ranges of equilibrium, the CNM and the WNM with
temperatures $\sim 50$ and $\sim 8000$ K, separated by a region of
unstable temperatures (Wolfire et al 1995; WHMTB).  Stable thermal
pressure equilibrium, and thermal pressure equality between the phases,
is a cornerstone of the MO theory. 

	Observational temperatures are derived in different ways for the
CNM and WNM.  For the CNM, the most accurate temperatures come from
comparing 21-cm line absorption and emission profiles.  The line opacity
$\propto {1 \over T}$; only the CNM produces significant absorption
lines, while the CNM and WNM both contribute to emission.  In contrast,
most WNM temperatures are derived from linewidths.  21-cm linewidths
provide upper limits to temperature, while in some cases combining them
with heavy element linewidths from optical/UV spectra provides actual
temperatures (see review by Heiles 2000). 

	Kulkarni \& Heiles (1987; KH) summarized temperature
measurements of the CNM and WNM, confining their analysis to 21-cm line
absorption/emission measurements because very few UV data were available
at that time.  For the CNM, they discussed single-dish results in terms
of the conventional interpretation of clouds within which the
temperature increases outwards.  As briefly discussed below, each
derived temperature is the lowest one in its cloud.  Expressed as
histograms, these coldest-cloud derived temperatures are broadly
distributed over the range $20 \rightarrow 300$ K (Mebold et al 1982,
MWKG; Dickey, Salpeter, \& Terzian 1978, DST; Payne, Salpeter, \&
Terzian 1982, PST).  In addition, there is a weakly significant
statistical relationship between the derived CNM temperature and the
21-cm line opacity, called the $T-\tau$ relation.  For the WNM, KH found
that the limited data supported a lower limit $\sim 5000$ K, but they
cautioned that the result needed confirmation. 

	In contrast, interferometric maps of a field around 3C147
(Kalberla, Schwarz, \& Goss 1985, KSG) provide a completely different
picture. For the CNM, the components are roughly isothermal and have
colder temperatures ($34 \rightarrow 74$ K). For the WNM, which
contributes $\sim 80\%$ of the mass, temperatures lie in the thermally
unstable range ($500 \rightarrow 2000$ K).  The high angular resolution
should make these results reliable.  However, such
results are available in only a few fields.  The fact that they conflict
with the single-dish results is disturbing.  The single-dish results are
the ones always quoted in reviews because of their much larger
statistical sample for the CNM; it is worth noting, though, that for the
WNM the single-dish sample is no larger than the interferometric one. 

	Here we present a statistical summary of new single-dish
temperatures.  For the CNM they are derived from HI absorption/emission
line data; for the WNM they are upper limits based on linewidths. We
introduce a new analysis technique for absorption/emission observations.
 Our results are consistent with the 3C147 results.  For the CNM, we
find lower temperatures than previous single-dish workers; these are
more in line with theoretical prediction for the CNM.  For the WNM we
find that most of the gas WNM is at thermally unstable temperatures;
this disagrees with the MO theory. 

\section{ New 21-cm Line Observations \label{radiotemps}}

\subsection{ Observations and reduction technique}

        Heiles and Troland (2001) are performing a survey of Zeeman
splitting of HI absorption lines with the Arecibo telescope\footnote{The
Arecibo Observatory is part of the National Astronomy and Ionosphere
Center, which is operated by Cornell University under a cooperative
agreement with the National Science Foundation.}.  These data have long
integration times, which produces excellent signal/noise and makes them
unsurpassed for obtaining temperatures.  Our results are more accurate
and cover more sources than the best previous surveys, which are the
Arecibo work by DST\footnote{We find about half the DST absorption
profiles to exhibit large differences from ours, with DST components
being wider and multiply-peaked; some of DST's profiles were corrupted
by local oscillator stability problems (Dickey, private communication). 
This completely explains the disagreement of Greisen and Liszt (1986)
with DST for 3C348.} and PST, and the Bonn/NRAO work by MWKG. Here we
report on 24 sightlines, 19 of which have $|b| > 20^\circ$ but otherwise
are randomly selected within Arecibo's declination range $\sim 0^\circ
\rightarrow 39^\circ$.

	Each absorption spectrum consists of very obvious velocity
components and we represent their optical depths by a set of $N$
Gaussians. Thus we least squares fit the observed spectrum ${T_{abs}
\over T_C} = e^{- \tau}$, where  
\begin{equation}
\tau = \Sigma_0^{N-1} \tau_{0n}
e^{-[(V-V_{0n})/ \delta V_n]^2} \ ;
\end{equation}
here $T_{abs} \over T_{C}$ is the
absorption profile divided by the continuum source strength and
$(V_{0n}, \delta V_n)$ are (central velocity, $1 \over e$ width) of
component $n$. We assume that each component is an independent physical
entity, and is {\it isothermal}. This is consistent with the findings of KSG. 

	Each HI emission spectrum contains structure but, also, is wider
than its associated absorption spectrum, as is well known. We assume
that the emission spectrum $T_{em} = T_{CNM} + T_{WNM}$, where 
$T_{CNM}$ is the contribution from the aforementioned CNM components.
$T_{WNM}$ is the contribution from $K$ additional wide Gaussians to
represent the WNM; $K$ is small number and often just one. The spin
temperatures in these additional components are so high that they have
negligible optical depth and produce no easily discernible features in
the absorption spectrum. 

	In least-squares fitting the emission, we include the absorption
of more distant CNM Gaussians by less distant ones.  Letting $T_n$ be
the spin temperature of component $n$, which is also the kinetic
temperature, 
\begin{equation}
T_{CNM} = \Sigma_0^{N-1} T_n (1 - e^{-\tau_n}) e^{-\Sigma_0^{M-1}\tau_m}
\ ,
\end{equation}
where the subscript $m$ with its associated optical depth
profile $\tau_m$ represents each of the $M$ CNM clouds that lie in front
of cloud $n$. For multiple absorption components, we experiment with all
possible orders along the line of sight and choose the one that yields
the smallest residuals.  We also include the absorption of each WNM
component by the CNM by assuming that a fraction $\mathcal F_k$ lies in
front of all the CNM and is unabsorbed, with the rest all lying behind;
thus
\begin{equation}
T_{WNM} = \Sigma_0^{K-1} [{\mathcal F_k} + (1-{\mathcal F_k})e^{-\tau}]  
  T_{0k} e^{-[(V-V_{0k}/ \delta V_k]^2} \ , \end{equation} where the
subscript $k$ represents each WNM component.  Note that $T_{0k}$ is a
brightness temperature, not a kinetic temperature.  In most cases
${\mathcal F_k}$ is indeterminant and we can only distinguish between
the two extremes ${\mathcal F_k} = (0,1)$.  The differences for
different orderings are sometimes not statistically significant but
nevertheless lead to differences in the derived CNM temperatures.  These
differences reflect the uncertainties in $T_n$ more than the
conventional errors derived from least squares fits.  Additional
uncertainties can occur if there are unresolved subcomponents.  We defer
discussion of these details to the more comprehensive paper (Heiles and
Troland 2001). 

	Previous single-dish authors, in contrast, implicitly assume
that clouds are {\it not} isothermal.  They derive the spin temperature
of a cloud at the peak of its absorption profile.  Thus each point on
their histograms represents the lowest derived temperature for that
particular cloud.  This temperature, however, is not the coldest
temperature in the cloud, because the line of sight also passes through
warmer gas. 
 
\subsection{ Sample result: 3C18}

\begin{figure}
\includegraphics{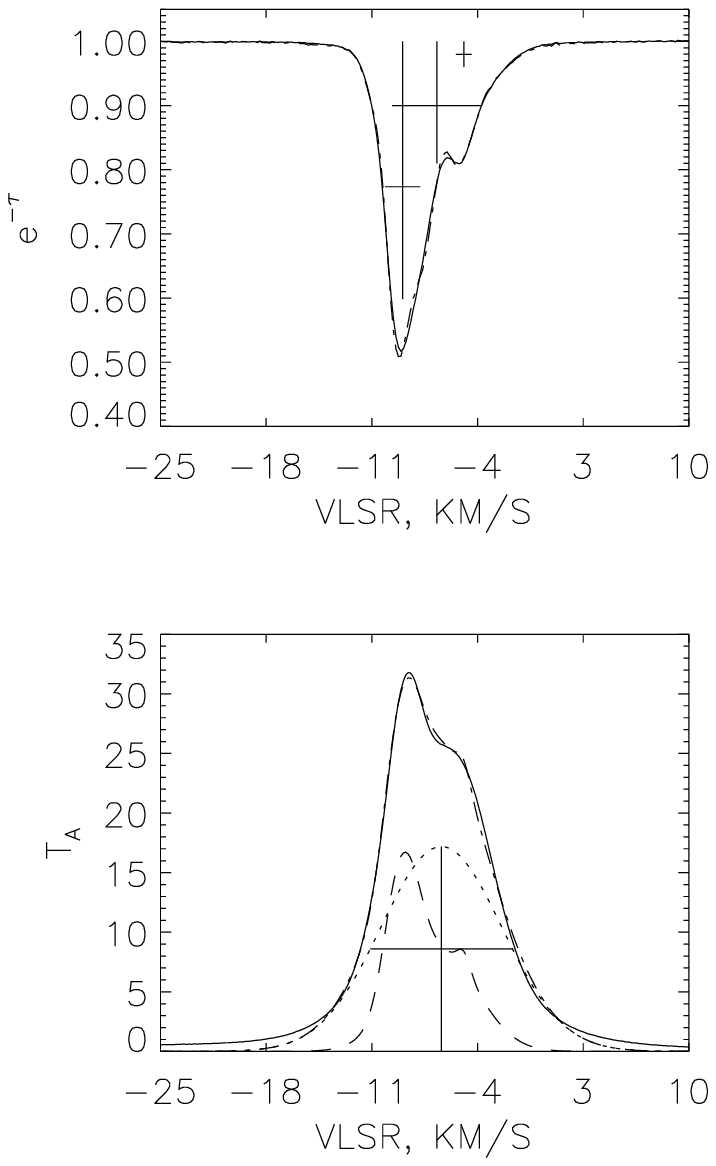}
\caption{ 21-cm line absorption (top) and emission (bottom) spectra for
3C18.  In the absorption spectrum, the solid line is data and the
dash-dot line the fit; crosses indicate the central $e^{-\tau}$'s and
halfwidths of the three Gaussian component parameters.  In the emission
spectrum, the solid line is the data; the dashed line is the
contribution from the three CNM components, the dotted line from
the WNM component, and the dash-dot line their sum. Dash-dot lines in
both figures are the fits and are so close to the data that they are
hard to distinguish. \label{3C18fig}} 
\end{figure}

	Figure~\ref{3C18fig} exhibits the results for 3C18 [located at
$(\ell, b) = (119^\circ, -53^\circ)$], which is a simple profile and
good for an illustrative example.  In the top panel the solid line is
the observed absorption spectrum $T_{abs} \over T_C$, which we fit with
the three CNM components whose depths and halfwidths are indicated; the
dash-dot line is the fit. In the bottom panel the solid line is the
observed emission spectrum $T_{em}$. The dashed curve is $T_{CNM}$; the
dotted is $T_{WNM}$ fit with $K=1$, which is unabsorbed by the CNM
because the lowest residuals are obtained with ${\mathcal F} = 1$. The
full fitted curve is the sum, shown as dash-dot, which is a good fit
except in the extreme line wings where stray radiation makes the data
suspect (e.g. Hartmann \& Burton 1997). 

	For 3C18, the WNM component has halfwidth 10.0 km s$^{-1}$,
which corresponds to purely thermal broadening at $T = 2200$ K; this is
an upper limit on the kinetic temperature $T_K$.  For the three CNM
components, left-to-right on Figure~\ref{3C18fig}, the halfwidths are 
$(2.5 \pm 0.03, 6.3 \pm 0.07, 1.1 \pm 0.15)$ km s$^{-1}$ and spin
temperatures are $32 \pm 1$, $43 \pm 6$, and $46 \pm 9$ K. The ratios of
total linewidth to thermal linewidth are $( 2.04 \pm 0.07, 4.50 \pm
0.63, 0.75 \pm 0.18)$; the $1\sigma$ uncertainty on the last ratio is
statistically consistent with a ratio $\ge 1$, as it must be.  The (WNM,
CNM) components contribute $N(HI) = (3.2, 1.8) \times 10^{20}$
cm$^{-2}$, respectively.  The WNM/CNM ratio is $\sim 1.8$, which is
close to our global average.
	
\subsection{ Our ensemble of WNM temperatures}

	Our 49 WNM components have linewidths that correspond to upper
limits on the kinetic temperature.  The top two panels of
Figure~\ref{histogramsfig} exhibit histograms of these limits, one for
number of components and one for column density.  Not included on these
histograms is one absorption component for which the spin temperature
was derived: $T_{spin} = 725$ K, $N(HI) = 1.4 \times 10^{20}$ cm$^{-2}$. 
Including this, 20 of the WNM components ($40\%$) have $T_K = 500
\rightarrow 5000$ K.  These contain $>47\%$ of the total WNM column
density; this is a lower limit because the WNM temperatures are upper
limits.  Because these components are not visible in absorption, their
spin temperatures exceed $\sim 500$ K.  This range, $500 \rightarrow
5000$ K, is approximately the thermally unstable range that separates CNM
from WNM. 

\subsection{ Our ensemble of CNM temperatures}

	Our 86 CNM temperatures are derived from absorption/emission
data and are values, not upper or lower limits.  The bottom two panels
of Figure~\ref{histogramsfig} exhibit histograms of these temperatures. 
The two histograms exhibit broad peaks in the range $T = 25 \rightarrow
75$ K.  47 of the CNM components ($54\%$) have temperatures in this
range; these contain $61\%$ of the total CNM column density.  We also
see colder gas: 10 components ($11\%$) containing $5\%$ of the mass have
$T = 10 \rightarrow 25$ K. We discount the four small-$N(HI)$ components
having $T < 10$ K: they are weak and the temperatures have large
errors. We find no support for the weakly significant $T-\tau$ relation
reviewed by KH.

\subsection{ Ratio of CNM and WNM components and mass}

	For WNM gas ($T > 500$ K), we found 49 components with total
$N(HI) = 107 \times 10^{20}$ cm$^{-2}$.  For CNM gas ($T < 200$ K) we
found 80 components having a total $N(HI) = 65 \times 10^{20}$
cm$^{-2}$.  In the mildly ambiguous range $T = 200 \rightarrow 500$ K we
found 6 components with total $N(HI) = 7.5 \times 10^{20}$ cm$^{-2}$. 
Thus, the ratio of CNM to WNM is, in terms of number of components, 1.6;
in terms of mass, 0.60.  Overall, our results indicate that about $60\%$
of all the neutral atomic ISM is WNM, with $T > 500$ K. 

\subsection{ Temperatures from optical/UV absorption line observations
\label{uvtemps}}

	To derive kinetic temperatures from an atomic optical/UV
absorption line, one decomposes it into Gaussian components.  Then one
does the same with the 21-cm emission line toward the star, fixing the
central velocities to be the same.  The comparison of line widths
separates the thermal and turbulent broadening.  The derived
temperatures are upper limits because the HI line comes from a much
larger angular area so the nonthermal component of its width may be
larger than that of the heavy element lines.  Such temperatures are
probably the best one can do for the WNM, but for CNM gas they are much
less accurate than those derived from 21-cm absorption/emission line
data. 

	Spitzer \& Fitzpatrick (1995) and Fitzpatrick \& Spitzer (1997)
use this technique towards two high-latitude stars and derive
temperatures and column densities for 21 diffuse neutral components.  Of
this total, 3 components have $T > 5000$ K, 13 have $T < 500$ K, and 5
have $T$ in the unstable $500 \rightarrow 5000$ K range; thus, $24\%$ of
the components are thermally unstable; these contain $63\%$ of the mass.

	One can derive excitation temperatures of the low-J states of
H$_2$ using UV absorption lines (Shull et al 2000; Spitzer, Cochran, \&
Hirshfeld 1974).  These tend to agree with previous 21-cm line
temperatures and are systematically higher than ours.  If our
temperatures are correct, then the low-J states are nonthermally
populated, as are the high-J ones.

\section{DISCUSSION AND COMPARISON WITH THEORY}

\subsection{ The WNM}

	Both our new HI and the optical/UV observations show that much
of the WNM---at least $47\%$---lies at temperatures that are unstable to
isobaric perturbations.  Our Arecibo data show this departure from
thermal stability in a statistically convincing manner.  Previous 21-cm
line studies have hinted at this result.  In emission/absorption
studies, MWKG decomposed emission line profiles into Gaussians, with
similar results; however, they didn't explicitly point out this
departure.  Verschuur \& Magnani (1994) and Heiles (1989) analyzed
emission profiles and found numerous components with widths in this
range, but without absorption data could not conclusively state that the
kinetic temperatures were indeed so high. 

	The large fraction of WNM in the thermally unstable regime
violates a fundamental cornerstone of equilibrium ISM models such as MO,
which all rely on thermal pressure equilibrium to push the gas into one
of the thermally stable CNM or WNM phases.  This result seems to push us
towards other types of model.  Two possibilities include time-dependent
models such as the supernova-dominant model of Gerola, Kafatos, and
McCray (1974) and turbulence-dominated models such as
V\'azquez-Semadeni, Gazol, \& Scalo (2000). 

\subsection{ The CNM}

\begin{figure}
\includegraphics{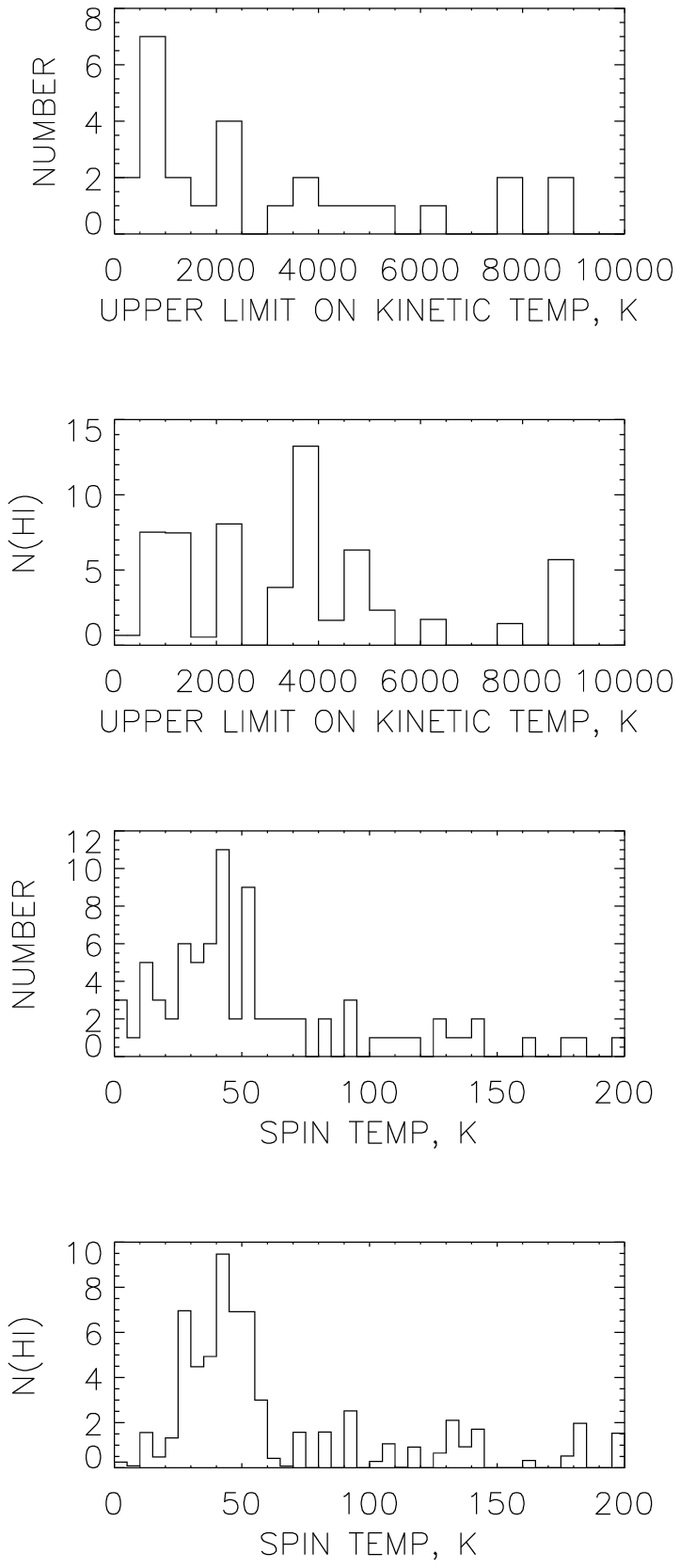}
\caption{ Histograms of derived temperatures.  The top two frames are
upper limits on kinetic temperature for the WNM derived from line
widths; of these, the top gives the number of Gaussian components and
the bottom the HI column density in units of $10^{20}$ cm$^{-2}$.  Off
the graphs to the right, with upper limits exceeding $10^4$ K,  are 21
components ($44\%$) containing $N(HI)=45 \times 10^{20}$ cm$^{-2}$ 
($43\%$).  The bottom two frames are values (not limits) of spin
temperature derived from absorption/emission data.  Off the graphs to
the right, with spin temperatures exceeding $200$ K, are 8 components
($9\%$)containing $N(HI)=37 \times 10^{20}$ cm$^{-2}$ ($36\%$). 
\label{histogramsfig}} 
\end{figure}

	Figure~\ref{histogramsfig} exhibits the histogram of derived
spin temperatures for all CNM components.  Both most of the components
and most of the mass have $T = 25 \rightarrow 75$ K.  This is in marked
contrast to previous results, where histograms were broad over the
ranges $20 \rightarrow 140$ K (MWKG) and $50 \rightarrow 300$ K (DST,
PST).  Our range is narrower and, moreover, temperatures extend to very
low values, with significant contributions down to $T=10$ K. 

	The peak above $T \sim 25$ K agrees very well with  the high
angular resolution results of KSG and, also, theory.  WHMTB included all
known processes in calculating their standard model, for which the CNM
equilibrium temperatures range from $25 \rightarrow 200$ K (the
corresponding densities are $n_{HI} \ga 1000 \rightarrow 4$ cm$^{-3}$). 
Our observed temperature range is smaller and corresponds to $n_{HI} \ga
250 \rightarrow 20$ cm$^{-3}$ and ${P \over k} = 10000 \rightarrow 1500$
cm$^{-3}$ K.  These numbers are in accord with ISM pressure measurements
(Jenkins, Jura, \& Lowenstein 1983).  

	Temperatures as low as our $10 \rightarrow 20$ K range can occur
in the absence of the PAH-type grains that produce grain heating (WHMTB;
Bakes and Tielens 1994).  In this case, heating is by photoionization of
Carbon and cooling by electron recombination onto ionized Carbon
(Spitzer 1978).  Such cold (and even colder) gas was invoked by Heiles
(1997) to help understand the existence of tiny-scale atomic structure;
the present results are encouraging for that interpretation. 

\acknowledgments

	It is a pleasure to acknowledge conversations with John Dickey,
Alex Lazarian, Jeff Linsky, Chris McKee, Paolo Padoan, Tom Troland, 
Enrique V\'azquez-Semadeni, and Ellen Zweibel.  This work was partly
supported by NSF grant AST9530590 to the author.

\clearpage

\end{document}